\documentclass{mn2e}
\usepackage{graphicx,epsfig}
\usepackage{floatflt}
\usepackage {amssymb}
\usepackage {amsmath}
\usepackage {longtable}
\title[Gravitational potential of a torus]
    {Gravitational potential of a homogeneous circular torus: \\ new approach}
\author[Bannikova, Vakulik, $\&$ Shulga]
  {E.Yu.~Bannikova$^{1,2}$, V.G.~Vakulik$^{1,2}$, V.M.~Shulga$^1$, \\
   $1$   Institute of Radio Astronomy of Nat.Ac.Sci. of Ukraine, Krasnoznamennaya 4,
   61022 Kharkov, Ukraine \\
   $^2$  Karazin Kharkov National University,
   Sumskaya 35, 61022 Kharkov, Ukraine \\
   E-mail: bannikova@astron.kharkov.ua, vakulik@astron.kharkov.ua, shulga@rian.kharkov.ua}

\date{Accepted 2010 September 13.  Received 2010 September 10; in
original form 2010 June 4}

\pagerange{\pageref{firstpage}--\pageref{lastpage}} \pubyear{2010}

\def\LaTeX{L\kern-.36em\raise.3ex\hbox{a}\kern-.15em
    T\kern-.1667em\lower.7ex\hbox{E}\kern-.125emX}

\begin{document}

\label{firstpage}

\maketitle

\begin{abstract}
The integral expression for gravitational potential of a
homogeneous circular torus composed of infinitely thin rings is
obtained. Approximate expressions for torus potential in the outer
and inner regions are found. In the outer region a torus potential
is shown to be approximately equal to that of an infinitely thin
ring of the same mass; it is valid up to the surface of the torus.
It is shown in a first approximation, that the inner potential of
the torus (inside a torus body) is a quadratic function of
coordinates. The method of sewing together the inner and outer
potentials is proposed. This method provided a continuous
approximate solution for the potential and its derivatives,
working throughout the region.
\end{abstract}


\section{Introduction}
Toroidal structures are now detected in astrophysical objects of
various types. Such objects are, for example, the ring galaxies,
where a ring of stars is observed. In some galaxies, the ring-like
distribution of stars is believed to be due to collisions of
galaxies, as, for example, in M31 (Block et al.\ 1987), and Arp
147 (Gerber et al.\ 1992). The analysis of the SDSS data  (Ibata
et al.\ 2003) indicates the existence of a star ring in the Milky
Way on scales of about 15-20 kpc, which is believed to be
originated from the capture of a dwarf galaxy. Obscuring tori are
observed in central regions of active galactic nuclei (AGN) (Jaffe
et al.\ 2004) and play an essential role in the unified scheme
(Antonucci 1993; Urry \& Padovani 1995). Ring-like structures
exhibit themselves in dark matter as well. An example can be the
galaxy cluster C10024+17 where a ring-like structure has been
found in distribution of dark matter with the use of gravitational
lensing method (Jee et al.\ 2007). In the Milky Way, the rotation
curves together with the EGRET data can be explained by  existence
of two rings of dark matter located at distances of about  4 kpc
and 14 kpc from the Galaxy center (de Boer et al.\ 2005). Such
toroidal structures can possess a significant mass, and thus
 gravitationally affect the matter motion.

B. Riemann devoted one of his last works to the gravitational
potential of a homogeneous torus (see Collected papers, 1948).
This work remained unfinished however. For over a century, no
attention has been paid to a torus gravitational potential
\footnote{In electrostatics, a potential of conducting torus shell
is used (Smythe 1950) that is much easier than the case when torus
density is uniform inside its volume.}. Kondratyev (2003) has
returned to this problem for the first time. In this work an exact
expression for the potential of a homogeneous torus on the axis of
symmetry was obtained. In (Kondratyev 2007) the integral
expressions for a homogeneous torus potential were found using a
disk as a primordial gravitating element. Stacking up such disks
will result in a torus with the potential equal to a sum of
potentials of component disks. However,  it is evident that any
integral expressions are problematic to use both in analytic
studies and in numerical integration of motion equations, and also
in solving the problems of gravitational lensing. B.Kondratyev et
al. (2009, 2010) have obtained an expansion of torus potential in
Laplace series, but showed, however, that such an expansion is
impossible inside some spherical shell.

In this paper we propose a new approach to investigation of the
gravitational potential of a torus. Special attention has been
paid to finding approximate expressions for the potential, which
would simplify investigation of astrophysical objects with
gravitating tori as structural elements. In contrast to
(Kondratyev 2007), we used an infinitely thin ring as a torus
component. Such ring is actually a realization of a torus, with
its minor radius tending to zero and the major one equaling the
ring radius. Using such an approach, we obtained an integral
expression for the potential of a homogeneous circular torus
(Section 2) and approximate expressions for the potential in the
outer (Section 3) and inner (Section 4) regions. In Section 5, the
method of determining a torus potential for the entire region is
suggested.

\section{Gravitational potential of a homogeneous torus}

Compose a torus with mass $M$, outer (major) radius $R$ and minor radius
$R_0$, of a set of infinitely thin rings - component rings hereafter,
- (see Fig.~1), with their planes being parallel to the torus symmetry
plane.
\begin{figure}
 \includegraphics[width = 84mm]{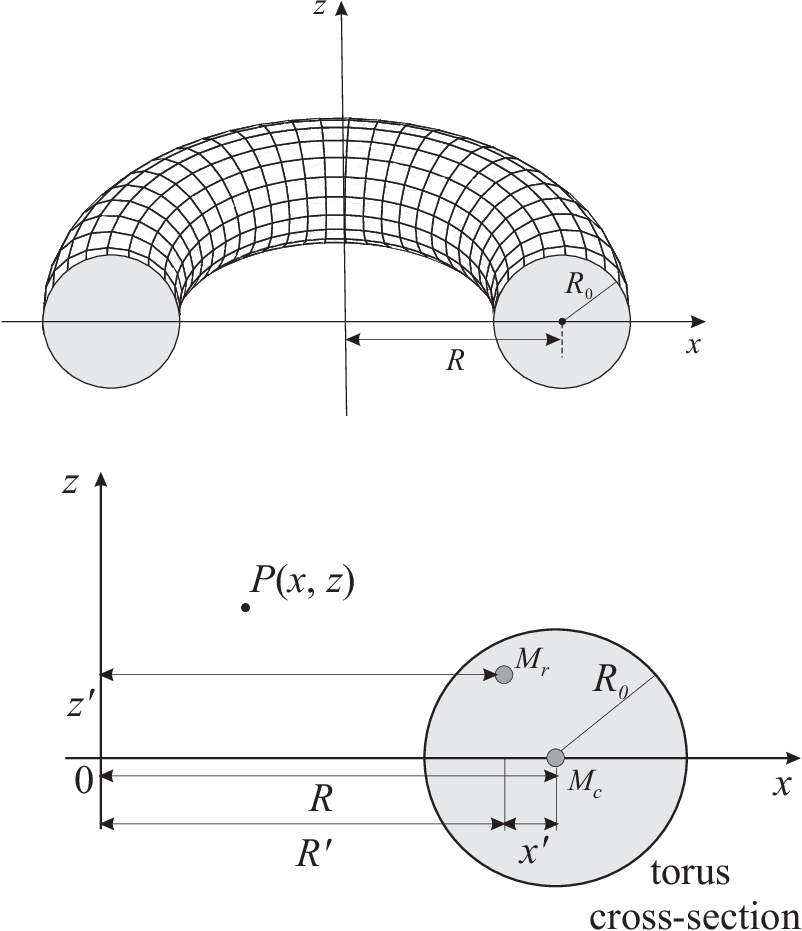}
 \caption{On the top: a 3D scheme of a torus configuration; on the bottom:
 schematic cross-section of a torus as a set of infinitely thin rings.}
\end{figure}
Select a central ring with mass $M_c$ and radius $R$ from a set
of rings composing a torus. The potential produced by this ring at
an arbitrary point $P(x,z)$ has a form:
\begin{equation}\label{eq_2.1}
  \varphi_c(x,z;R,M_c)=\frac{G M_c}{\pi R}\cdot
  \phi_c\left( \frac{x}{R},\frac{z}{R}\right)
\end{equation}
where the dimensionless potential of the infinitely thin ring is
\begin{equation}\label{eq_2.2}
    \phi_c\left( \frac{x}{R},\frac{z}{R}\right) =
    \sqrt{ \frac{R}{x}\cdot m}\cdot K(m),
\end{equation}
$K(m)=\int_0^{\pi/2}\frac{d\beta}{\sqrt{1-m \sin^2 \beta}}$ is the
complete elliptical integral of the first kind, and its parameter
is
\begin{equation}\label{eq_2.3}
  m=\frac{4xR}{(x+R)^2+z^2}.
\end{equation}
The potential at a point $P(x,z)$ produced by an arbitrary ring
with radius $R'$ and mass $M_r$ located in the torus at a hight
$z'$ (Fig.~1), has a form
\begin{equation}\label{eq_2.4}
  \varphi_r(x,z;M_r,z') = \frac{GM}{\pi R'}\cdot \phi_r
\end{equation}
where the expression for  $\phi_r$ is obtained by substitution
in (\ref{eq_2.2}) of a kind $x/R \rightarrow x/R'$ and $z/R
\rightarrow (z-z')/R'$. Denote a ring coordinate \footnote{Here,
the coordinates of a point where the ring intersects the plane of the
torus cross-section are meant as the ring coordinates. Position of
this point is determined by the ring radius $R'$ and by the ring
distance from the torus plane of symmetry $z'$.} counted off the
center of a torus cross-section (Fig.~1) by $x'=R'-R$, and
therefore, its radius is $R'=R+x'$. We may conveniently introduce
the dimensionless coordinates $\eta' = x'/R$, $\zeta' = z'/R$ and
$\rho = x/R$, $\zeta = z/R$ that will result in an expression for
the dimensionless potential of the form
\begin{equation}\label{eq_2.5}
  \phi_r(\rho,\zeta;\eta',\zeta') = \sqrt{\frac{(1+\eta')\cdot
  m_r}{\rho}}\cdot K(m_r)
\end{equation}
where
\begin{equation}\label{eq_2.6}
  m_r = \frac{4\rho\cdot (1+\eta')}{(1+\eta'+\rho)^2 +
  (\zeta-\zeta')^2}~.
\end{equation}
From a condition of the torus homogeneity, mass-to-radius ratios
for the central and arbitrary component rings are the same, and
thus, $M_c = M_r/(1+\eta')$. Expression for the potential of the
component ring is then
\begin{equation}\label{eq_2.7}
\varphi_r(\rho,\zeta;\eta',\zeta') = \frac{GM_c}{\pi R}\cdot
\phi_r
\end{equation}
where $\phi_r$ is determined by expressions (\ref{eq_2.5}),
(\ref{eq_2.6}). Due to additivity, the torus potential can be
represented as the integral over potentials of the component
rings. To do this, we replace a discrete mass of the ring $M_c$ in
(\ref{eq_2.7}) by a differential $dM$, which for the homogeneous
torus equals $dM = \frac{M}{\pi r_0^2}d\eta'd\zeta'$, where $M$ is
a total mass of the torus equaling to a sum of masses of the
component rings, and $r_0 = R_0/R$ is a dimensionless minor radius
of the torus (a geometrical parameter). Then, the potential of the
homogeneous circular torus takes the form
\begin{equation}\label{eq_2.8}
  \varphi_{torus}(\rho,\zeta) = \frac{GM}{\pi^2 R
  r_0^2}\int_{-r_0}^{r_0} \int_{-\sqrt{r_0^2 - \eta'^2}}^{\sqrt{r_0^2 - \eta'^2}}
  \phi_r(\rho,\zeta;\eta',\zeta')d\eta'd\zeta'
\end{equation}
This integral expression for the torus potential is valid for both
the inner and outer points. The validity of expression
(\ref{eq_2.8}) is confirmed by calculation of the potential made
by direct integration over the torus volume.
\begin{figure}
 \includegraphics[width = 84mm]{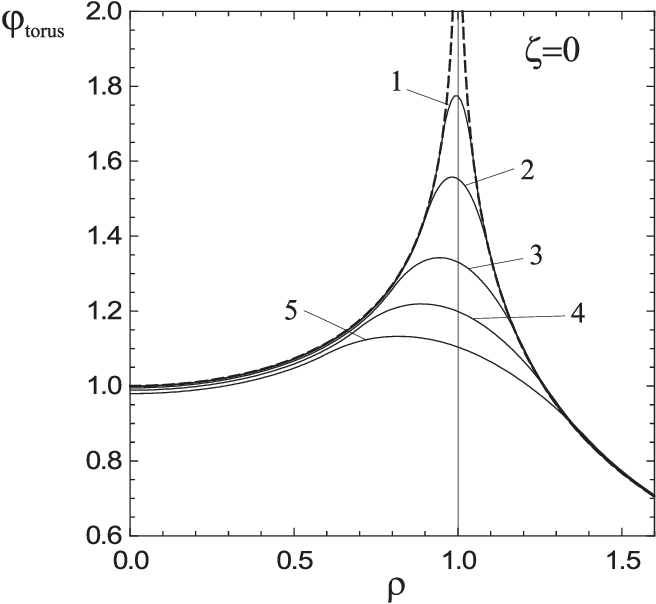}
 \caption{Dependence of the potential on the radial coordinate
 for $\zeta = 0$, for the tori with various values of geometrical parameter:
 1) $r_0=0.05$, 2) $r_0=0.1$, 3) $r_0=0.2$, 4) $r_0=0.3$,
 5) $r_0=0.4$. The potential of an infinitely thin ring (\ref{eq_2.1}) with mass
 $M$ equal to the mass of the torus is shown by the dashed line. In this plot
 and in all the subsequent figures, $M=1$, $R=1$, $G =1$.}
\end{figure}
Hereafter, in analyzing approximate expressions, we will use the
term "exact" $~$ for the values of potential obtained from the
integral formula (\ref{eq_2.8}). In Fig.~2, dependencies of the
torus potential on the radial coordinate are presented, which were
obtained numerically from formula (\ref{eq_2.8}) for tori with
different values of the geometrical parameter $r_0$. The potential
curves for all values of $r_0$ are seen to be inscribed into the
potential curve of the infinitely thin ring of the same mass and
radius, located in the torus symmetry plane. The potential curve
to the right of the torus surface ($\rho > 1 + r_0$) virtually
coincides with the potential curve of the ring, while to the left
($\rho < 1 + r_0$), it passes lower and differs by a quantity that
depends on $r_0$ (see section 3). In Fig.~3 the dependencies of the torus
potential on the radial coordinate are presented, which were calculated from
expression (\ref{eq_2.8}) for different values of $\zeta$.
\begin{figure}
 \includegraphics[width = 84mm]{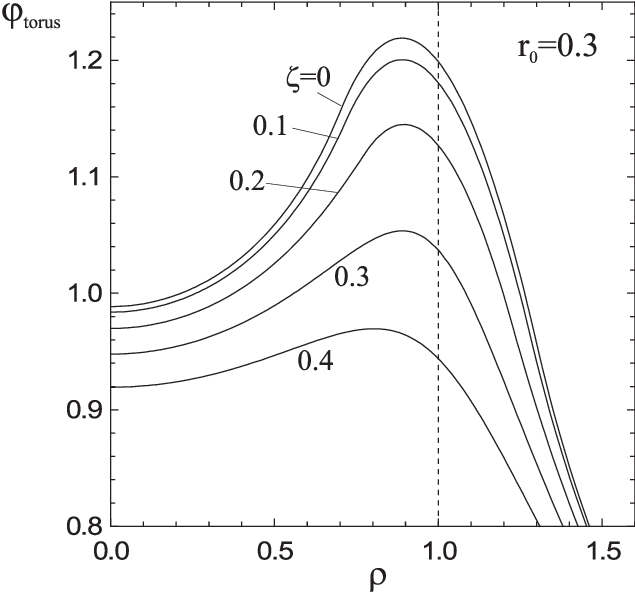}
 \caption{Potential of the torus with $r_0 = 0.3$ as a function of
 the radial coordinate for various values of $\zeta = 0, 0.1, 0.2,
  0.3, 0.4$.}
 \end{figure}

Note, that in contrast to the work by Kondratyev (2007, p.196,
expression (7.26)), where the torus potential is expressed only
through a single integration of the elliptical integrals of all
the three kinds, the torus potential (\ref{eq_2.8}) in our work is
expressed by double integration of the elliptical integral of the
first kind. However, further analysis of this expression
 for the torus potential (\ref{eq_2.8}) allows us to obtain
approximations that are physically understandable  and enable
solving practical astrophysical tasks which need multiple
calculations of the gravitational potential of the torus.

For further analysis of the torus potential,  we define the inner
region as the volume bounded by the torus surface (inside the
torus body) and outer region as the region outside this surface.

\section{Torus potential in the outer region}

It is seen from Fig.~2 that the outer potential of the torus can
be approximately represented by the potential of an infinitely
thin ring of the same mass up to torus surface. For
$\rho\rightarrow 0$, the values of the torus potential and
potential of the infinitely thin ring differ by a quantity that
depends on a geometric parameter $r_0$, that is especially evident
for a thick torus ($r_0 > 0.5$). Find a relationship between the
outer potential of the torus and the potential of a ring of the
same mass, that is, derive an approximate expression for torus
potential in the outer region, where a condition $(\rho - 1)^2 +
\zeta^2\geq r_0^2$ holds. Within this region, the integrand
$\phi_r(\rho,\zeta;\eta',\zeta')$ in (\ref{eq_2.8}) does not have
singularities for all $\eta'$,$\zeta'$, therefore, it can be
expanded as the Maclaurin series in powers of $\eta'$,$\zeta'$ in
the vicinity of a point $\eta'=\zeta'=0$. Since the integrals in
symmetrical limits from the series terms that contain cross
derivatives and derivatives of the odd orders are equal to zero,
only summands with the even orders remain in the expansion. With
the quadratic terms of the series being restricted, the potential
of the component ring is:
\begin{equation}\label{eq_3.1}
 \phi_r(\rho,\zeta;\eta',\zeta') \approx
 \phi_c(\rho,\zeta)+\frac{1}{2}\left.
                     \frac{\partial^2\phi_r}{\partial\eta'^2}
                                \right|_{\substack{\eta'=0 \\
                                                   \zeta'=0}}
                                \eta'^2
                  + \frac{1}{2}\left.
                     \frac{\partial^2\phi_r}{\partial  \zeta'^2}
                                \right|_{\substack{\eta'=0 \\
                                                    \zeta'=0}}
                                \zeta'^2.
\end{equation}
Substituting (\ref{eq_3.1}) into (\ref{eq_2.8}), we will have
after integration:
\begin{equation}\label{eq_3.2}
  \varphi_{torus} \approx \frac{GM}{\pi R}\phi_c \cdot
        \left(
          1 + \frac{r_0^2}{2\phi_c}
                \left[
                     \left.
                \frac{\partial^2\phi_r}{\partial  \eta'^2}
                     \right|_{\substack{\eta'=0 \\
                                        \zeta'=0}} +
                     \left.
                \frac{\partial^2\phi_r}{\partial  \zeta'^2}
                     \right|_{\substack{\eta'=0 \\
                                        \zeta'=0}}
                \right]
        \right).
\end{equation}
Ultimately, the approximate expression for the torus potential in
the outer region $(\rho - 1)^2 + \zeta^2 \geq r_0^2$ has a form:
\begin{equation}\label{eq_3.3}
   \varphi_{torus}(\rho,\zeta; r_0) \approx \frac{GM}{\pi R}\phi_c \cdot
        \left(
         1 - \frac{r_0^2}{16} + \frac{r_0^2}{16} \cdot
         S(\rho,\zeta)
        \right),
\end{equation}
where $\phi_c = \sqrt{\frac{m}{\rho}}K(m)$  is a dimensionless
potential of the central ring (\ref{eq_2.2}), and
\begin{equation}\label{eq_3.4}
  S(\rho,\zeta) = \frac{\rho^2 + \zeta^2 - 1}{(\rho - 1)^2 + \zeta^2}
                    \cdot \frac{E(m)}{K(m)},
\end{equation}
$E(m) = \int_0^{\pi/2}d\beta\sqrt{1-m \sin^2\beta}$ is  the
complete elliptical integral of the second kind. We may
conveniently proceed to a new variable $\eta = \rho - 1$ that
allows expression (\ref{eq_3.4}) to be represented as
\begin{equation}\label{eq_3.5}
  S(\eta,\zeta) = \frac{\eta^2 + \zeta^2 + 2\eta}{\eta^2 + \zeta^2}
  \cdot \frac{E(m)}{K(m)},
\end{equation}
where
\[
m=4\frac{\eta + 1}{(\eta +2)^2 + \zeta^2}.
\]
Expression (\ref{eq_3.3}) for the torus potential (we will further
call it the S-approximation), with (\ref{eq_3.4}) or
(\ref{eq_3.5}) taken into account, represents the torus potential
accurately enough in the outer region $\eta^2 + \zeta^2 \geq
r_0^2$ (Fig.~2). Since $\mid S\mid\leq 1$ the second multiplier in
(\ref{eq_3.3}) is a slowly varying function in $\rho$ and $\zeta$.
Let us simplify the expression (\ref{eq_3.3}) replacing the second
multiplier by its asymptotic approximations.

In the first case,  $\rho \rightarrow 0$ corresponding to $\eta
\rightarrow -1$, the parameter $m \rightarrow 0$ and $E(m)/K(m)
\rightarrow 1$, therefore, $S \rightarrow (\zeta^2 - 1)/(\zeta^2 +
1)$. The expression for the torus potential in this case is
\begin{equation}\label{eq_3.6}
  \varphi_{torus}(\rho,\zeta;r_0) \approx \frac{GM}{\pi R}
  \phi_c(\rho,\zeta)\cdot \left(
               1 - \frac{r_0^2}{16}+\frac{r_0^2}{16}\frac{\zeta^2 - 1}
                                                         {\zeta^2 + 1}
                          \right).
\end{equation}
Since the dimensionless potential of the ring (\ref{eq_2.2}) at
the symmetry axis is $\phi_c = \frac{\pi}{\sqrt{1 + \zeta^2}}$, we
get for the torus
\begin{equation}\label{eq_3.7}
  \varphi_{torus}(0,\zeta;r_0) \approx \frac{GM}{R}
  \frac{1}{\sqrt{1 + \zeta^2}} \cdot \left(
               1 - \frac{r_0^2}{16}+\frac{r_0^2}{16}\frac{\zeta^2 - 1}{\zeta^2 + 1}
                          \right).
\end{equation}
and for $\zeta=0$,
\begin{equation}\label{eq_3.8}
  \varphi_{torus}(0,0;r_0) \approx \frac{GM}{R}\left(
                                1 - \frac{r_0^2}{8}
                                                \right).
\end{equation}
The second \textbf{summand} $GM/R \cdot r_0^2/8$ in (\ref{eq_3.8})
describes displacement of the torus potential at the symmetry axis
as compared to the potential of an infinitely thin ring (Fig.~2).

In the second case, at large $\eta$, the parameter $m \rightarrow
0$, and $S \rightarrow 1$  in (\ref{eq_3.3}), therefore:
\begin{equation}\label{eq_3.9}
  \varphi_{torus}(\rho,\zeta;r_0) \approx \frac{GM}{\pi R}
  \phi_c(\rho,\zeta),
\end{equation}
that is, the torus potential is equal to the potential of the
infinitely thin ring with the same mass $M$ and radius $R$ in this
case.
\begin{figure}
 \includegraphics[width = 84mm]{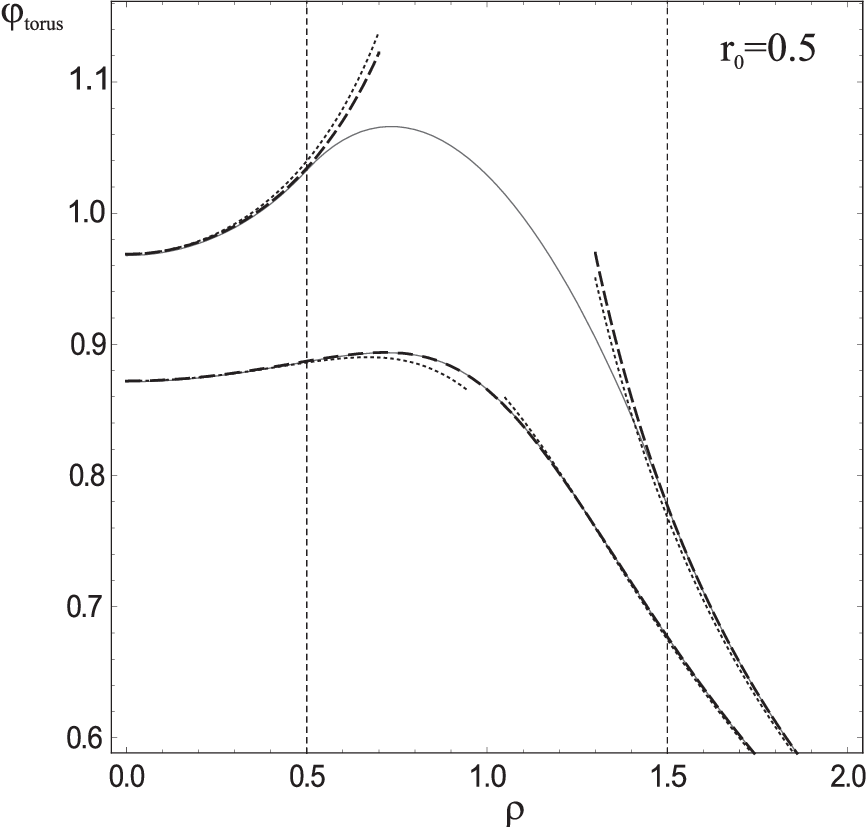}
 \caption{Dependence of the torus potential with $r_0= 0.5$ on $\rho$ for
 $\zeta=0$ (upper curves) and $\zeta=0.5$ (lower curves). Solid lines show
 the potential  calculated with the exact formula (\ref{eq_2.8}). The
 S-approximation (\ref{eq_3.3}) of  the potential is shown in dashed lines,
 and dotted lines represent  the limiting cases of the S-approximation: the
 potential curve for  the infinitely thin ring (\ref{eq_3.9}) is to the right of the
 torus cross-section, and the curve representing a "shifted" $~$ potential of
 the infinitely thin ring (\ref{eq_3.6}) is to the left. The boundaries of the torus
 cross-section are dotted with the vertical lines.}
 \end{figure}
 It is seen from Fig.~4 that {\it the $S$-approximation for the torus outer
 potential (\ref{eq_3.3})  is applicable up to the torus surface}$~$(upper
 curves). Indeed, in the region
 $\rho \leq 1 - r_0$, difference between the potential obtained from
 the integral expression (\ref{eq_2.8}) and its value taken from
 the S-approximation reaches maximum near the torus surface and
 does not exceed $0.2\%$ for $r_0=0.5$. The difference remains small
 even for a thick torus: it does not exceed $1.5\%$ for $r_0=0.9$.
 For $\zeta = r_0$ all the points are outer, and the curves for
 the exact potential and $S$-approximation virtually coincide
 (deviation is less than $0.1\%$).

Note that asymptotics of the $S$-approximation for the outer
potential (\ref{eq_3.7}) and (\ref{eq_3.9}) also describe the
torus potential well enough (dotted line in Fig.~4). Thus, for
$|\zeta| < r_0$, the approximation (\ref{eq_3.7}) can be used to
estimate the potential inside the region bounded by a cylinder
with radius $\rho - r_0$, while the approximation (\ref{eq_3.9})
is applicable outside the region bounded by a cylinder with radius
$\rho + r_0$. At $|\zeta| \gg 1$, expression (\ref{eq_3.7}) tends
to (\ref{eq_3.9}), and expression for potential of the infinitely
thin ring (\ref{eq_2.2}) can be used within the whole outer region
to approximately evaluate the torus potential.

Therefore, \textit{ the outer potential of the torus can be
represented with good accuracy by a potential of an infinitely
thin ring of the same mass.  The dependence of the geometrical
parameter $r_0$ appears only in the torus hole; it is taken into
account in the "shifted"$~$potential of the infinitely thin ring
(\ref{eq_3.6}). These approximations are valid up to the surface
of the torus. \footnote{There is some analogy with the known
result: the outer potential of a solid sphere of mass $M$ is the
same as that generated by a point mass $M$ located at the sphere's
center. Note, however, that torus has another system of
equigravitating elements (Kondratyev, 2007).}}

\section{Torus potential in the inner region}

To analyze the inner potential of the torus, it is convenient to
select the origin of a coordinate system in the center of the
torus cross-section (Fig.~5). Then, the dimensionless potential of
the central ring takes a form:
\begin{equation}\label{eq_4.1}
  \phi_c(\eta,\zeta) = \sqrt{\frac{m}{1 + \eta}}\cdot K(m)
  \end{equation}
where
\[
m = \frac{4(1 + \eta)}{(2 + \eta)^2 +  \zeta^2}.
\]
\begin{figure}
\centering
 \includegraphics[width = 64mm]{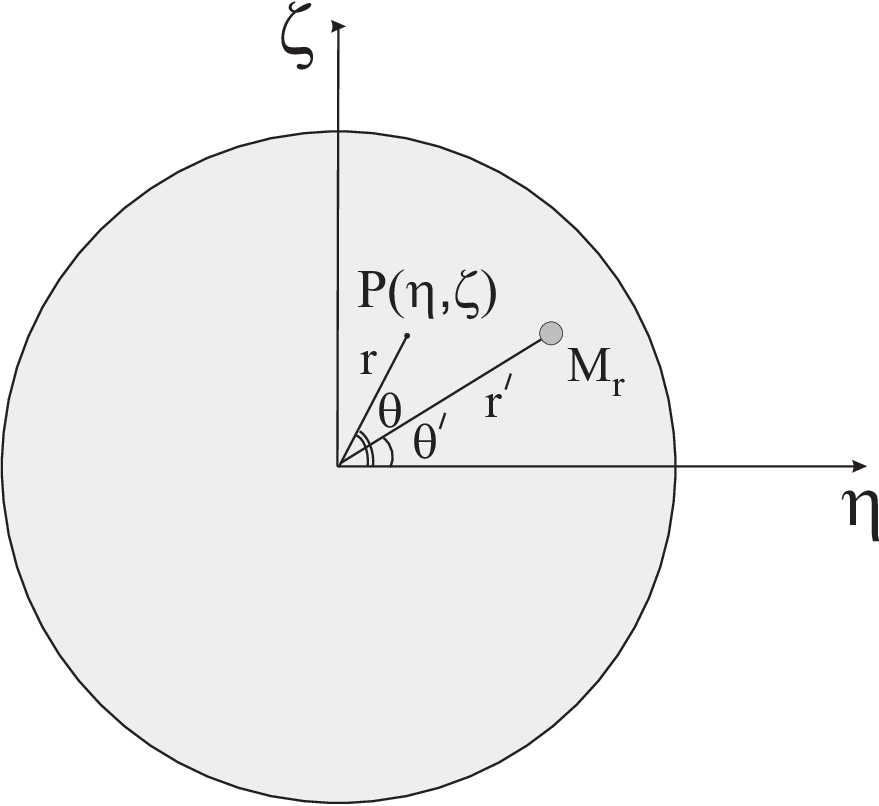}
 \caption{The scheme of torus cross-section.}
 \end{figure}
Consider the potential of the central ring (\ref{eq_4.1}) in the
vicinity of $\eta \rightarrow 0$, $\zeta \rightarrow 0$, that
corresponds to $m \rightarrow 1$. In this case, the elliptical
integral in (\ref{eq_4.1}) can be expanded in terms of a small
parameter $m_1 = 1-m$. With the series clipped by two terms, we
will have:
\begin{equation}\label{eq_4.2}
  K(m_1) \approx \ln \frac{4}{\sqrt{m_1}} +
                 \frac{1}{4}m_1 \ln\frac{4}{e\sqrt{m_1}},
\end{equation}
where
\[
m_1 = \frac{r^2}{r^2 + 4(1 + \eta)},
\]
$r^2 = \eta^2 + \zeta^2$. The approximate formula for the ring
potential expressed through the parameter $m_1$ is then:
\begin{equation}\label{eq_4.3}
  \phi_c(\eta,\zeta)\approx\frac{2\sqrt{m_1}}{r}
            \left(
            \ln \frac{4}{\sqrt{m_1}} + \frac{1}{4}m_1\ln \frac{4}{e\sqrt{m_1}}
            \right).
\end{equation}
Passage to the potential of an arbitrary component ring is
fulfilled by substitutions $1 + \eta \rightarrow (1 + \eta)/(1 +
\eta')$ and $\zeta \rightarrow (\zeta - \zeta')/(1 + \eta')$,
which results in an expression
\begin{equation}\label{eq_4.4}
  \phi_r(\eta,\zeta;\eta',\zeta') \approx \frac{2(1 + \eta')}{\sqrt{q}}
                \left[
                \ln\frac{4}{\sqrt{m'_1}}\left(1 + \frac{m'_1}{4}\right)
                - \frac{m'_1}{4}
                \right]
\end{equation}
where $m'_1 = (\mathbf{r} - \mathbf{r}')^2/q$, $q = (\mathbf{r} -
\mathbf{r}')^2 +4(1 + \eta)(1 + \eta')$. A summand $(\mathbf{r} -
\mathbf{r}')^2 = (\eta - \eta')^2 + (\zeta - \zeta')^2$  is a
square of the distance between the component ring and a point $P$
(Fig.~5). Expansion (\ref{eq_4.4}) is valid for $m'_1 \rightarrow
0 $, therefore, $(\mathbf{r} - \mathbf{r}')^2 \ll 1$.

Confine ourselves by the case of a thin torus ($r_0 \ll 1$ ). Then
$(\mathbf{r} - \mathbf{r}')^2 \ll 4(1 + \eta)(1 + \eta')$ and $(1
+ \eta)(1 + \eta') \approx 1$, and the first multiplier in
(\ref{eq_4.4}) can be written to the second-order terms as:
\begin{equation}\label{eq_4.5}
  f_1 \equiv \frac{2(1 + \eta')}{\sqrt{q}} \approx
  \sqrt{\frac{1 + \eta'}{1 + \eta}}
      \left(
      1 - \frac{1}{8}(\mathbf{r} - \mathbf{r}')^2
      \right).
\end{equation}
After expanding the square root in (\ref{eq_4.5}) in powers of
$\eta$ and $\eta'$, we obtain
\begin{equation}\label{eq_4.6}
   f_1 \approx
   \left(
      1 + \frac{\eta'}{2} - \frac{\eta}{2} - \frac{\eta \eta'}{4}
      - \frac{\eta'^2}{8} + \frac{3}{8}\eta^2
   \right)
   \left(
      1 - \frac{1}{8}(\mathbf{r} - \mathbf{r}')^2
      \right).
\end{equation}
 Similarly, the second multiplier (in square brackets) in
expression (\ref{eq_4.4}) can be written to the terms quadratic in
coordinates as:
\[
f_2 \equiv \ln \frac{4}{\sqrt{m'_1}}
   \left(
      1 + \frac{m'_1}{4} \right)
      - \frac{m'_1}{4}
    \approx \frac{1}{2}(\eta + \eta') - \frac{1}{4}(\eta^2 +
    \eta'^2)+
\]
\begin{equation}\label{eq_4.7}
      + \ln\frac{8}{\mid\mathbf{r} - \mathbf{r}'\mid} +
    \frac{(\mathbf{r} - \mathbf{r}')^2}{16}
    \ln\frac{8e}{\mid\mathbf{r} - \mathbf{r}'\mid}.
\end{equation}
Thus, we obtain the following expression for the potential of the
component ring:
\begin{equation}\label{eq_4.8}
  \phi_r(\eta,\zeta;\eta',\zeta') \approx f_1\cdot f_2.
\end{equation}
In consideration of the inner potential of the torus, rewrite
expression (\ref{eq_2.8}) in the polar coordinates (Fig.~5):
\begin{equation}\label{eq_4.9}
  \varphi_{torus}(r,\theta;r_0) = \frac{GM}{\pi^2 R r_0^2}
        \int_0^{r_0} \int_0^{2\pi} \phi_r(r,\theta;r',\theta')
        r'dr'd\theta',
\end{equation}
where coordinates of the component ring are $\eta' =
r'\cos\theta'$, $\zeta' = r'\sin\theta'$, and coordinates of a
point $P$ are $\eta = r\cos\theta$, $\zeta = r\sin\theta$.
Substitute  (\ref{eq_4.6}) and (\ref{eq_4.7}) into (\ref{eq_4.8}),
and after multiplying, restrict ourselves by the terms quadratic
in $\eta$, $\zeta$ and $\eta'$, $\zeta'$. Then, after integration
of (\ref{eq_4.9}), we obtain the approximate expression for the
inner potential of the torus:
\begin{equation}\label{eq_4.10}
  \varphi_{torus}(\eta,\zeta;r_0) \approx \frac{GM}{2\pi R}
        \left[
           c + \tilde{a}_1\eta + \tilde{a}_2\eta^2 + \tilde{b}_2\zeta^2
        \right],
\end{equation}
where
\[
k \equiv \frac{r_0}{8}, \quad c = 1 + 2k^2 - 2\ln k + 8k^2\ln k,
\quad \tilde{a}_1 = 1 + \ln k,
\]
\[
\tilde{a}_2 = -\frac{1}{(8k)^2} - \frac{1}{16}(11 + 10\ln k), \]
\[
\tilde{b}_2 = -\frac{1}{(8k)^2} + \frac{1}{16}(3 + 2\ln k).
\]
The first summand in (\ref{eq_4.10}) is the value of the torus
potential in the center of the torus cross-section: $c =
\phi_{torus}(0,0;r_0)$. To further analyze the inner potential, it
is convenient to transfer to a coordinate system normalized to the
geometrical parameter of the torus $r_0$. Then the series
coefficients will transform to the form:
\[
a_1 = 8k(1 + \ln k), \qquad a_2 = -1 - 4k^2(11 + 10\ln k),
\]
\[
b_2 = -1 + 4k^2(3 + 2\ln k).
\]
The expression for the torus potential (\ref{eq_4.10}) can be
written as
\begin{equation}\label{eq_4.11}
  \varphi_{torus}(\eta,\zeta;r_0) \approx
  \frac{GM}{2\pi R} \left[
         c + a_1 \frac{\eta}{r_0} + a_2
         \left(\frac{\eta}{r_0}\right)^2 +b_2
            \left(\frac{\zeta}{r_0}\right)^2
                    \right].
\end{equation}
It follows from (\ref{eq_4.11}) that the maximal value of the
potential reaches at a point $\eta_{max} = -(a_1r_0)/(2a_2)$,
$\zeta = 0$, while equipotential lines are ellipses with their
centers displaced an amount $\eta_{max}$ with respect to the
center of the torus cross-section, and a ratio of semiaxes of the
ellipses is $\sqrt{b_2/a_2}$. Note, that location of the potential
maximum $\eta = \eta_{max}$, $\zeta = 0$ corresponds to the
weightlessness point, where the resultant of all the forces
affecting a particle inside the torus equals zero. In such an
approximation, components of the force inside the torus depend on
the coordinates linearly at that. In Fig.~6, the curves of the
inner potential in the coordinate system normalized to $r_0$ are
presented for three values of $r_0$.
\begin{figure}
 \includegraphics[width = 84mm]{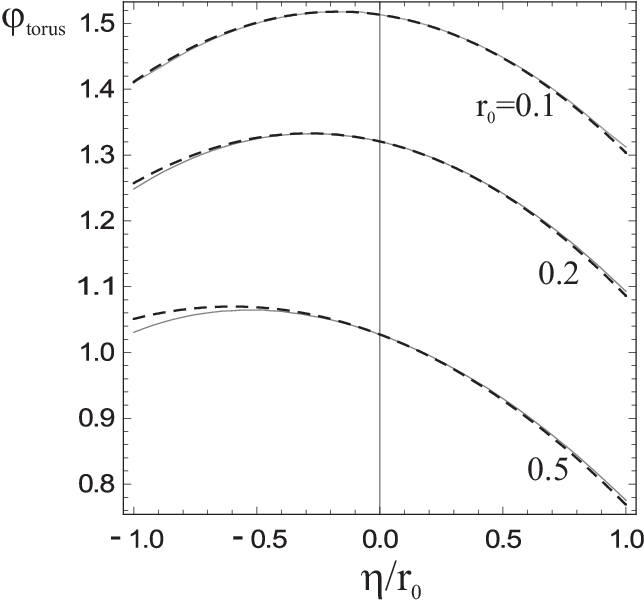}
 \caption{Dependence of the inner potential on the normalized coordinate
  $\eta/r_0$ at $\zeta = 0$ for various values of the geometric parameter:
   $r_0 = 0.1, 0.2, 0.5$. Solid curves represent dependencies of the potential
   on the distance to the center of the torus cross-section, which were calculated
   with the exact formula (\ref{eq_2.8}). Dashed lines are the potential curves
   taken with the approximate expression for the inner potential
   of the torus (\ref{eq_4.11}).}
 \end{figure}

Though we confined ourselves to the case of a thin torus, the
curves of potential taken from expression (\ref{eq_4.11}) are well
consistent with the curves for the exact potential (\ref{eq_2.8})
up to $r_0 = 0.5$, where the deviation is maximal near the torus
surface and is of the order of  $2\%$ (Fig.~6). The value of
potential in the center of the torus cross-section (a constant $c
= \phi_{torus}(0,0;r_0)$ in (\ref{eq_4.10})), also coincides with
its exact value.

It is of interest to investigate the solutions obtained for
limiting cases. Indeed, the case $r_0 = R_0/R \rightarrow 0$
corresponds to two limiting passages to an infinitely thin ring
($R$ is fixed while $R_0 \rightarrow 0$) and to a cylinder, when
$R_0$ is fixed and $R \rightarrow \infty$. Dwell on the limiting
passage to the cylinder potential. At $r_0 \rightarrow 0$, the
coefficients $c \rightarrow 1 - 2\ln k$, $a_1 \rightarrow 0$, and
coefficients $a_2,b_2 \rightarrow -1$, and expression
(\ref{eq_4.11}) takes the following form in this case:
\begin{equation}\label{eq_4.12}
  \varphi_{torus}(\eta,\zeta;r_0 \rightarrow 0)
  \approx \frac{GM}{2\pi R}
  \left[
     1 - 2\ln\frac{r_0}{8} - \left(\frac{r}{r_0}\right)^2
  \right]
\end{equation}
where $r^2 = \eta^2 + \zeta^2$. It is known that the inner
potential of a circular cylinder with the length $2H$, much larger
than the radius $R_0$ of its cross-section (Kondratyev 2007), has
the form:
\begin{equation}\label{eq_4.13}
  \varphi_{cyl} = \frac{GM}{2H}
      \left[
        2\ln\frac{2\sqrt{e}H}{R_0} -
                        \left(\frac{r}{r_0}\right)^2
      \right].
\end{equation}
After a formal substitution $2H = 2\pi R$ in (\ref{eq_4.13}), (the
cylinder length is equal to the length of the central ring), we
get an expression:
\begin{equation}\label{eq_4.14}
\varphi_{cyl} = \frac{GM}{2\pi R}
      \left[
       1 - 2\ln \left( \frac{r_0}{2 \pi}\right) -
                        \left(\frac{r}{r_0}\right)^2
      \right].
\end{equation}
Expression (\ref{eq_4.14}) coincides with (\ref{eq_4.12}) to a
constant \footnote{Difference in the constant may be caused by the
curvature of the  torus surface.}.

The quadratic dependence on $r$ for the inner potential of a thin
torus can be also derived in the case, when the minor radius $R_0
\rightarrow 0$. The outer potential of the torus was shown in
section~3 to be approximately equal to the potential of an
infinitely thin ring of the same mass and radius. In this case,
the smaller is the torus geometrical parameter $r_0$, the more
accurate is this approximation. Therefore, at $\eta^2 + \zeta^2
\geq r_0 \rightarrow 0$, the outer potential of the torus tends to
the potential of an infinitely thin ring.  In this case,
$\eta,\zeta \rightarrow 0$, and thus, the elliptical integral in
the expression for an infinitely thin ring (\ref{eq_2.2}) can be
expanded in the vicinity of $m  \rightarrow 1$. If we confine
ourselves to the first term of the expansion, we get an
approximate expression for the potential of a central infinitely
thin ring
\begin{equation}\label{eq_4.16}
  \varphi_c(\eta,\zeta) \approx \frac{GM}{2\pi R}
        \left(
          -\ln(\eta^2 + \zeta^2) + 2\ln8
        \right),
\end{equation}
which remains valid for the outer potential of the thin torus as
well. It should be noted that there is no dependence on $r_0$,
because in such an approximation, all thin tori with the same
masses and major radii are equigravitating for the outer
potential. The derivatives of the outer potential of the thin
torus in $\eta$, $\zeta$ are then:
\[
\frac{\partial \varphi_c}{\partial \eta} \approx
            -\frac{GM}{\pi R}\frac{\eta}{r^2},
\qquad \frac{\partial \varphi_c}{\partial \zeta} \approx
            -\frac{GM}{\pi R}\frac{\zeta}{r^2}
\]
and take the following forms at the torus surface
($\eta^2+\zeta^2=r_0^2$):
\[
 \left.\frac{\partial \varphi_c}{\partial \eta}\right|_{r=r_0} \approx
            -\frac{GM}{\pi R r_0}\cos \theta,
\qquad \left.\frac{\partial \varphi_c}{\partial
\zeta}\right|_{r=r_0} \approx
            -\frac{GM}{\pi R r_0}\sin \theta
\]
It is the linear dependence of the force on coordinates $\eta$,
$\zeta$ that satisfies such boundary conditions. Thus, the inner
potential of the thin torus can be represented to the integration
constant in the form:
\begin{equation}\label{eq_4.17}
  \varphi_{torus}(\eta,\zeta;r_0) \approx \frac{GM}{2\pi R}
        \left(
          -\frac{\eta^2 + \zeta^2}{r_0^2} + c(r_0)
        \right).
\end{equation}
Equating (\ref{eq_4.16}) with (\ref{eq_4.17}) at the torus
surface, we obtain the expression for the constant $c(r_0) =
-2\ln(r_0/8) + 1$ that coincides with expression (\ref{eq_4.10})
obtained above at $r_0 \ll 1$.

It becomes evident from analysis of the inner potential for the
two limiting cases ($R_0 \rightarrow 0$  and $R \rightarrow
\infty$) that the first summand in coefficients $a_2$, $b_2$ of
the power series (\ref{eq_4.11}) represents properties of the inner
potential of a cylinder. With the cylinder potential separated,
the inner potential of the torus (\ref{eq_4.11}) can be written
as:
\begin{equation}\label{eq_4.18}
  \varphi_{torus}(\eta, \zeta) = \varphi_{cyl}(r) +
  \varphi_{curv}(\eta, \zeta),
\end{equation}
where
\begin{equation}\label{eq_curv}
 \varphi_{curv} \approx \frac{GM}{2\pi R} \left[c_{curv} +
a_1 \left(\frac{\eta}{r_0}\right) + c_a
\left(\frac{\eta}{r_0}\right)^2 +
c_b\left(\frac{\zeta}{r_0}\right)^2\right],
\end{equation}
\[
c_{curv} = 2\ln\frac{8}{2\pi} + 2k^2(1 + 4\ln k),
\]
\[
c_a = 1 + a_2, \qquad c_b = 1 + b_2.
\]
The second summand $\varphi_{curv}(\eta,\zeta)$, which we will
call a potential of curvature, implies  curvature of the torus
surface. Indeed, all the coefficients of the series
(\ref{eq_curv}) tend to zero in the limiting passage to the
cylinder ($r_0 \rightarrow 0$), and $\varphi_{curv} \rightarrow
0$. Therefore, \textit{ the inner potential of the torus can be
represented as a sum of the cylinder potential and a term
comprising a geometrical curvature of the torus surface}.

\section{Sewing together the inner and outer potentials at the torus surface}

In the previous sections, we derived approximate expressions for
the torus potential in the outer ($\eta^2 + \zeta^2 \geq r_0^2$)
and inner ($\eta^2 + \zeta^2 \leq r_0^2$) regions. It has been
shown also that the inner potential of the torus can be
represented by a series in powers of $\eta / r_0$ and $\zeta /
r_0$, and the constant, linear and quadratic terms of the series
were determined analytically. To find a larger number of the
series terms sufficient to represent the inner potential
accurately enough, and to obtain a continuous approximate solution
for the potential and its derivatives in the whole region that
would satisfy the boundary conditions at the surface, we will act
in the following way. Represent the inner potential of the torus
as a power series \footnote{We consider a dimensionless potential
here. To pass to the dimensional case, (\ref{eq_5.1}) must be
multiplied by $GM/R$.}:
\[
  \phi(\eta,\zeta;r_0) = \frac{1}{2\pi}
    \left(
     c(r_0) + \sum_{i=1}a_i(r_0)
                            \left(
                            \frac{\eta}{r_0}
                            \right)^i + \right.
\]
\begin{equation}\label{eq_5.1}
            \left. + \sum_{i=1}\sum_{j=1}t_{ij}(r_0)
                            \left(\frac{\eta}{r_0}\right)^i
                            \left(\frac{\zeta}{r_0}\right)^j +
             \sum_{j=1}b_j(r_0)
                            \left(
                            \frac{\zeta}{r_0}
                            \right)^j
    \right),
\end{equation}
where $c(r_0)$, $a_i(r_0)$, $b_j(r_0)$, $t_{ij}(r_0)$ are unknown
coefficients. Note, that the series (\ref{eq_5.1}) contains only
the terms with even powers of $\zeta$, because the torus potential
is symmetric in $\zeta$.

Suppose that we have an analytical expression for the torus
potential $\Psi(\eta,\zeta;r_0)$ at its surface ($\eta^2 + \zeta^2
= r_0^2$). Also, write down the inner potential of the torus
(\ref{eq_5.1}) at its surface ($\eta = r_0\cos \theta$ and $\zeta
= r_0\sin \theta$):
\[
\phi(\theta, r_0) = \frac{1}{2\pi}
    \left(
     c + \sum_{i=1} a_i \cos^i \theta \right. +
\]
\begin{equation}\label{eq_5.2}
     \left. + \sum_{i=1}\sum_{j=1}t_{ij}
                            \cos^i \theta
                            \sin^j \theta +
             \sum_{j=1}b_j  \sin^j \theta
    \right).
\end{equation}
From conditions of equality of the inner and outer potential and
its derivatives in coordinates at the torus surface for several
angles $\theta_k$, we obtain a system of $3k$ linear equations to
determine coefficients $c$, $a_i$, $b_j$, $t_{ij}$:
\begin{equation}\label{eq_5.3}
\left\{
\begin{array}{l}
 c + \sum\limits_{i = 1} a_i  \cos^i \theta _k
        + \sum\limits_{i = 1} \sum\limits_{j = 1} t_{ij} \cos^i \theta _k  \sin^j \theta _k + \\
        + \sum\limits_{j = 1} b_j   \sin^j \theta _k
                = 2\pi \,\Psi (\theta _k ,r_0 ) \\\\
 \sum\limits_{i = 1} i\cdot a_i \cos^{i-1} \theta _k  +
 \sum\limits_{i = 1} \sum\limits_{j = 1} i \cdot t_{ij} \cos^{i-1} \theta _k \sin^j \theta _k = \\
        = 2\pi r_0 \frac{\partial}{\partial \eta}
        \Psi (\theta _k ,r_0 )  \\\\
 \sum\limits_{j = 1} {j\cdot b_j}  \sin^{j-1} \theta _k  +
 \sum\limits_{i = 1} \sum\limits_{j = 1} j\cdot t_{ij} \cos^i \theta _k  \sin^{j-1} \theta _k = \\
         = 2\pi r_0 \frac{\partial }{\partial \zeta } \Psi (\theta _k ,r_0 )
 \end{array}
\right.
\end{equation}
Thus, if we had the analytic solution for the outer potential of
the torus, we could obtain an exact expression for the inner
potential as an infinite series in powers of $\cos\theta_k$,
$\sin\theta_k$, using the boundary conditions and solving the
system of equations (\ref{eq_5.3}). Since there is no analytic
expression for the outer potential, we can use the above
approximate expression (\ref{eq_3.3}) for the torus potential  in
the outer region (the S-approximation), and introduce
designations:
\[
\begin{array}{l}
\Phi = \sum\limits_k \left[
                    \phi_{in}(\theta_k,r_0) -
\phi_{out}(\theta_k,r_0)
                    \right]^2 \\
\Phi_1 = \sum\limits_k \left(\frac{\partial}{\partial\eta}
                \left[
                    \phi_{in}(\theta_k,r_0) -
\phi_{out}(\theta_k,r_0)
                 \right]
                    \right)^2 \\
\Phi_2 = \sum\limits_k \left(\frac{\partial}{\partial\zeta}
                \left[
                    \phi_{in}(\theta_k,r_0) -
\phi_{out}(\theta_k,r_0)
                 \right]
                    \right)^2,
\end{array}
\]
where $\phi_{in}$, $\phi_{out}$ are solutions for the inner
(\ref{eq_5.2}) and outer (\ref{eq_3.3}) potentials at the torus
boundary, respectively. The unknown coefficients of the series can
be then determined from a condition of the minimal value of a
functional:
\begin{equation}\label{eq_5.4}
  F = \Phi + \Phi_1 + \Phi_2 \rightarrow \text{min}.
\end{equation}
The functional (\ref{eq_5.4}) was minimized with the least squares
method, and coefficients of the series (\ref{eq_5.2}) were
determined up to the 4-th power. The coefficients of the series
are presented in Appendix (Table~A1). In Fig.~7 dependence of the
potential on the radial coordinate from the exact expression
(\ref{eq_2.8}) is presented for the entire region, as well as its
approximate solution obtained by sewing together the
S-approximation (\ref{eq_3.3}) and the inner potential
(\ref{eq_5.2}).
\begin{figure}
 \includegraphics[width = 84mm]{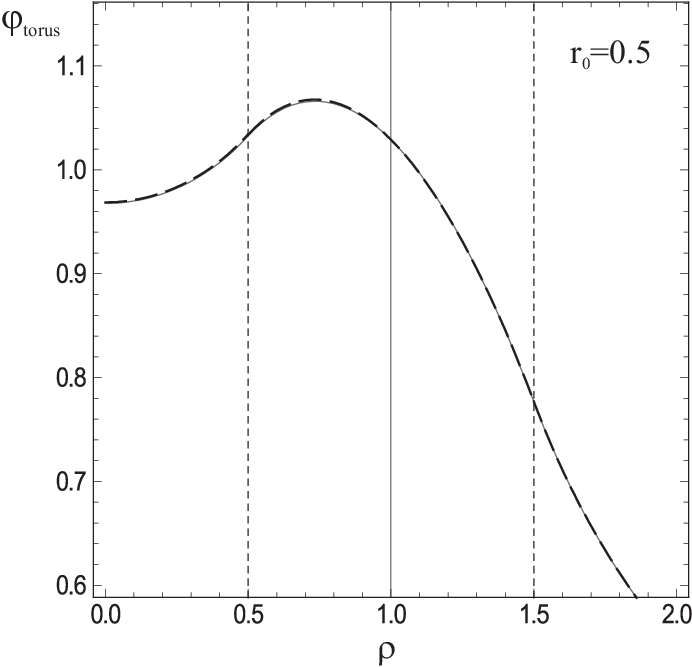}
 \caption{Dependence for the torus potential on $\rho$ for $r_0=0.5$ ($\zeta = 0$) in the whole region:
 the potential curve calculated from the exact expression (\ref{eq_2.8}) is shown by
 a solid line; dashed line demonstrates a result of sewing together the
 S-approximation in the outer region (\ref{eq_3.3}) with the inner potential represented
 by the power series (\ref{eq_5.2}) with the coefficients found from the sewing
 condition (\ref{eq_5.4}).}
 \end{figure}
Though the approximate solutions were obtained assuming that the
torus is thin $r_0 \ll 1$, we see that the exact (\ref{eq_2.8})
and approximate solutions are consistent even for the torus with
$r_0 = 0.5$.
\begin{figure}
 \includegraphics[width = 84mm]{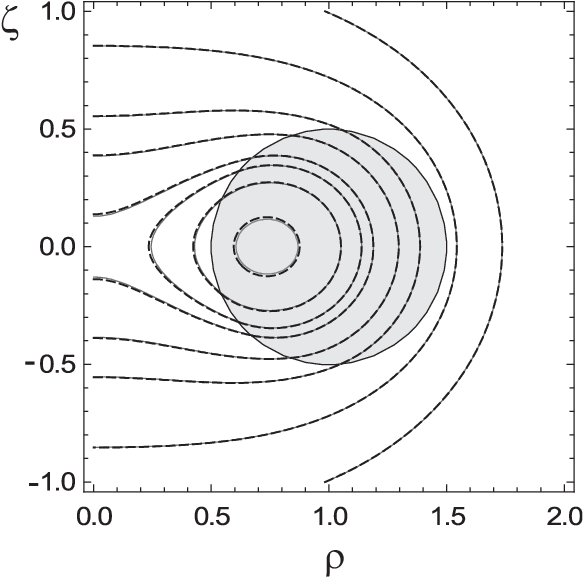}
 \caption{Equipotential curves for the torus with $r_0=0.5$.
 Solid lines are those calculated from the exact expression
 for the potential; dashed lines are those for the approximate
 formulas for the outer and inner potentials (see the figure~7 caption).
 The torus body cross-section is indicated as a gray circle.}
 \end{figure}
In Fig.~8, the equipotential curves on the plane of the torus
cross-section are shown, where a good agreement for all values of
$\rho$, $\zeta$ is seen as well.

\newpage

\section{Conclusions}
In the present work, the gravitational potential of a homogeneous
circular torus is investigated in details. An integral expression
for its potential that is valid for an arbitrary point is obtained
by composing the torus of infinitely thin rings. This approach has
made it possible to find an approximate expression for the outer
potential of the torus (S-approximation), that has a sufficiently
simple form. It is shown that the outer potential of the torus can
be represented with good accuracy by a potential of an infinitely
thin ring of the same mass.  The dependence of the geometrical
parameter $r_0$ appears only in the torus hole; it is taken into
account in the "shifted"$~$potential of the infinitely thin ring.
These approximations are valid up to the surface of the torus.

For the inner potential, an approximate expression is found in the
form of a power series to the second-order terms, where the
coefficients depend only on the geometric parameter $r_0$.
Expressions for the potential in the center of the torus
cross-section and for coordinates of the potential maximum are
obtained, and the limiting passage to a cylinder potential is
considered. It is shown that the inner potential of the torus can
be represented as a sum of the cylinder potential and a term
comprising a geometrical curvature of the torus surface. A method
for determining the torus potential over the whole region is
proposed that implies sewing together at the surface of the outer
potential (S-approximation) with the inner potential represented
by the power series. This method provided a continuous approximate
solution for the potential and its derivatives, working throughout
the region.

Surely, matter distribution within a torus is inhomogeneous in
actual astrophysical objects, and a torus cross-section may differ
from a circular one. Therefore, it is further interesting to
account for inhomogeneity of matter distribution inside a torus,
for difference of the torus cross-section from a circular form,
and so on.

\section*{Acknowledgments}

This work was partly supported by the National Program
"CosmoMicroPhysics".

We thank Professor V.M. Kontorovich for some helpful suggestions
and Dr V.S. Tsvetkova for critical reading of the original version
of the paper.

\appendix

\section[]{Coefficients of the power series for inner potential of a torus}

In Table~A1, coefficients of the power series (up to the 4-th
power) for the inner potential of the torus, are presented, which
were calculated from the sewing condition for the torus with
various values of the geometrical parameter $r_0$. The analytic
expression (\ref{eq_4.11})  was used to determine the zero-th
coefficient $c$ of the series.
\begin{table*}
 \centering
 \begin{minipage}{140mm}
  \caption{Coefficients of the power series for the inner potential
  of the torus for various values of $r_0$, obtained with the method of sewing.}
  \begin{tabular}{rrrrrrrrrr}
   Coeff.     &         &     & \multicolumn{4}{c}{Geometrical parameter $r_0$}  \\
          &    0.1   &     0.2  &     0.3  &    0.4   &    0.5   &    0.6   &    0.7   &    0.8   &    0.9   \\ 
 $a_1$    & -0.33798 & -0.53651 & -0.68154 & -0.79129 & -0.87439 & -0.93587 & -0.97906 & -1.00628 & -1.01928 \\
 $a_2$    & -0.98002 & -0.93543 & -0.87773 & -0.81171 & -0.74107 & -0.66865 & -0.59677 & -0.52739 & -0.46224 \\
 $b_2$    & -1.00411 & -1.01086 & -1.01970 & -1.02892 & -1.03759 & -1.04495 & -1.05030 & -1.05299 & -1.05237 \\
 $a_3$    & 0.02392  & 0.04364  &  0.05781 &  0.06608 & 0.06853  & 0.06550  & 0.05753  & 0.04525  & 0.02938  \\
 $t_{12}$ & 0.02550  & 0.05329  & 0.08404  &  0.11791 & 0.15454  & 0.19323  & 0.23295  & 0.27238  & 0.30991  \\
 $a_4$    & -0.00182 & -0.00785 & -0.01610 & -0.02576 & -0.03580 & -0.04535 & -0.05371 & -0.06036 & -0.06495 \\
 $b_4$    & 0.00061  & 0.00131  & 0.00308  &  0.00570 & 0.00922  & 0.01362  & 0.01880  & 0.02453  & 0.03045  \\
 $t_{22}$ & -0.00122 & -0.00812 & -0.01948 & -0.03681 & -0.06076 & -0.09157 & -0.12900 & -0.17213 & -0.21931
\end{tabular}
\end{minipage}
\end{table*}
In Fig.~A1, the linear ($a_1$) and quadratic ($a_2$, $b_2$)
coefficients of the power series as functions of $r_0$ obtained
analytically from (\ref{eq_4.10}) are shown by solid lines; the
dots are the proper values of these coefficients obtained from the
condition of sewing (see Table~A1).  The values of the analytic
coefficients are seen to coincide up to $r_0 = 0.5$ with their
values obtained independently with the method of sewing from (\ref{eq_5.4}).
\begin{figure}
 \includegraphics[width = 84mm]{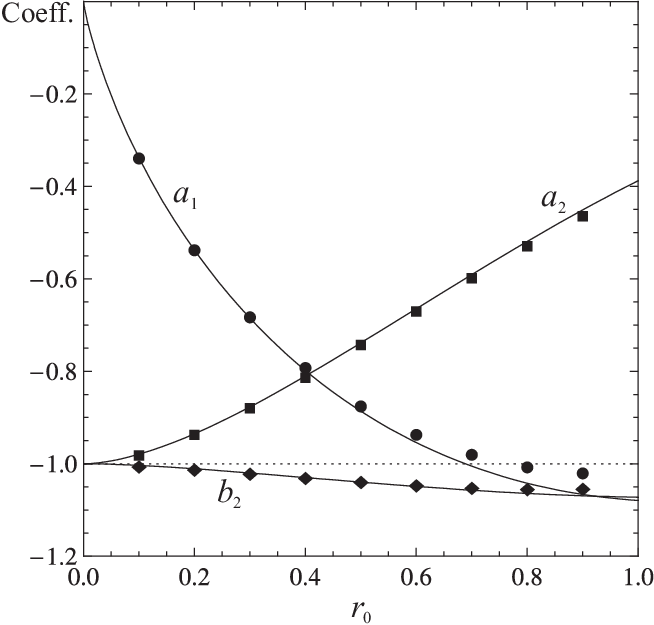}
 \caption{Dependence the first coefficients of
 the power series on $r_0$ of  for the inner potential obtained with the method of sewing.}
 \end{figure}
\end{document}